# VALIDASI DATA DENGAN MENGGUNAKAN OBJEK *LOOKUP* PADA BORLAND DELPHI 7.0

**Oleh: Leon Andretti Abdillah**
**Dosen Tetap Universitas Bina Darma**

***Abstracts***: *Developing an application with some tables must concern the validation of input (scpecially in Table Child). In order to maximize the accuracy and input data validation. Its called lookup (took data from other dataset). There are 2 (two) ways to lookup data from Table Parent: 1) Using Objects (DBLookupComboBox & DBLookupListBox), or 2) Arranging The Properties Of Fields Data Type (shown by using DBGrid). In this article is using Borland Delphi software (Inprise product). The methods is offered using 5 (five) practice steps: 1) Relational Database Scheme, 2) Form Design, 3) Object Databases Relationships Scheme, 4) Properties and Field Type Arrangement, and 5) Procedures. The result of this research are : 1) The relationship that using loopkup object is valid, and 2) Delphi's Lookup Object can be used for 1-1, 1-N, and M-N relationship.*

## 1. PENDAHULUAN

Dalam pembuatan suatu program aplikasi, sering kali suatu form tidak hanya terdiri atas sebuah tabel saja, melainkan dapat terdiri atas sejumlah tabel yang saling berhubungan. Untuk *field* tertentu (dari Tabel *Child*), datanya kadang-kadang membutuhkan data yang ada / bersumber dari suatu tabel/*dataset* yang sudah ada (Tabel *Parent*).

Seringkali pemasukan data (berupa *Foreign Key*) mengalami kendala, berupa: kurangnya validitas, ketidakkonsistenan data, kurang akurat, dll. Hal ini banyak terjadi pada mahasiswa yang sedang mengikuti ujian Skripsi/Tugas Akhir/PKL. Sehingga relasi antar tabel yang terlibat dalam aplikasi menjadi bias, tidak terintegrasi, dan menghasilkan *performance* yang cenderung rendah.

Penggunaan program generasi ke-empat dalam pembuatan suatu program aplikasi nyaris menjadi suatu standar minimal saat ini. Salah satu program yang berkembang dengan pesat, mudah digunakan, kinerja tinggi, adalah Borland Delphi (produk dari *Inprise Corp.*). Ia popular digunakan baik dari kalangan akademisi maupun praktisi, karena menggunakan objek bahasa dasar yaitu Bahasa Pascal yang *syntax*-nya sangat familiar dengan bahasa manusia (Bahasa Inggris). Delphi juga memiliki kekayaan dengan menyediakan suatu IDE (*Integrated Development Integrity*) yang membuat *programmer* nyaman dan memiliki sejumlah fasilitas dalam membangun programmnya, Delphi juga dikembangkan dengan konsep OOP (*Object Oriented Programming*) sehingga objek-objek yang ada dapat dipakai ulang (*re-use*).

Berdasarkan latar belakang di atas, penulis berminat untuk berbagi pengetahuan dalam meningkatkan validitas input data yang berupa *Foreign Key* pada suatu tabel, sehingga dapat meningkatkan integritas pembuatan program aplikasi.

Diharapkan karya ilmiah ini dapat digunakan baik oleh dosen, mahasiswa, maupun pembaca yang ingin menerapkan validasi dengan menggunakan *lookup* pada Borland Delphi.

## 2. TINJAUAN PUSTAKA

### Validasi

Validasi mengacu pada serangkaian aktivitas yang berbeda yang memastikan bahwa perangkat lunak yang dibangun dapat ditelusuri ke persyaratan pelanggan. Boehm dalam Pressman (2002:573).

Sehingga dapat diartikan validasi sebagai suatu cara untuk memastikan bahwa data yang kita masukkan untuk data dari suatu *field* (pada suatu tabel) adalah data yang benar/akurat/sesuai peruntukannya.

### Lookup

*If you have a Delphi form with data controls designed to allow editing the Applications table, you have to make sure that only TypeName values from the Types table can be applied to the Types field of the Application table. You also have to make sure that only AuthorName values from the Authors table can be applied to the Author field of the Application table. The best way to make this sure is to provide users with a string list to select the values from rather than having them enter values manually.*

*Both TDBLookupListBox and TDBLookupComboBox are data-aware controls that enable us to choose a value from another table or from a predefined list of values.* (www.About/Lookup!%20-%20DB%20Course-Chapter%2015%20-%20Page%201-4.htm)

(Jika Anda memiliki suatu form dengan data controls yang dirancang untuk membolehkan pengeditan tabel aplikasi, Anda harus memastikan bahwa hanya nilai TypeName dari tabel Type yang dapat diaplikasikan ke *field* Type dari tabel Authors. Anda juga harus memastikan bahwa hanya nilai AuthorName dari tabel Author yang dapat diaplikasikan ke *field* Author dari tabel Aplikasi. Cara terbaik untuk memastikannya adalah dengan menyediakan *users* dengan suatu *string list* untuk memilih nilai-nilai dibandingkan dengan memasukkan secara manual/langsung).

Baik TDBLookupListBox dan TDBLookupComboBox merupakan data-aware controls yang memungkinkan kita untuk memilih suatu nilai dari tabel lain atau dari daftar yang sudah tersedia (*predefined list*).

*Lookup* merupakan suatu istilah yang dapat digunakan untuk menggambarkan bagaimana suatu *field* data dari tabel/form tertentu berisi/mengambil data dari *dataset* lain. Sebagai ilustrasi anda dapat melihat skema *Relational Database* (RD) pada gambar 2.

Perhatikan tabel Parts (tbParts) merupakan tabel yang memiliki suatu *field* VendorNo. *Field* VendorNo tersebut sebenarnya merupakan *Primary Key* dari Tabel Vendors (tbVendors). Untuk meningkatkan tingkat akurasi pemasukan data VendorNo pada Tabel Parts, kita dapat menggunakan objek DBLookupComboBox/DBLookupListBox. Dengan cara ini maka sewaktu kita mengakses objek DBLookupComboBox/ListBox, kita seakan-akan mengkases isi dari Tabel Vendors.

Secara umum, *lookup* dapat dilakukan dengan 2 (dua) cara, yaitu : 1) Penggunaan Objek/Komponen *Lookup*, atau 2) Dengan Pengaturan *Properties*. Field Type. Objek *lookup* yang dapat digunakan adalah : 1) DBLookupComboBox, dan 2) DBLookupListBox.

**TDBLookupListBox (**  **)**

Merupakan objek/komponen pada Pages Data Controls yang digunakan untuk menyediakan *users* dengan suatu daftar *lookup* item yang nyaman untuk memasangkan suatu nilai *field* menggunakan nilai nilai-nilai dari suatu *field* dalam *dataset* lain. *Lookup* list boxes biasanya menampilkan nilai-nilai yang merepresentasikan deskripsi yang lebih bermakna dari suatu nilai *field* secara aktual. (Borland Delphi Help).

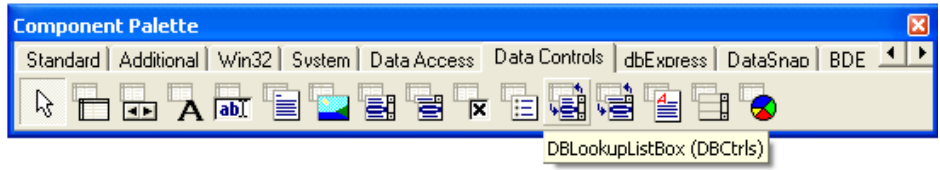

**Gambar 2.1 Objek DBLookupListBox**

**TDBLookupComboBox (** 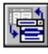 **)**

Merupakan objek/komponen pada Pages Data Controls yang digunakan untuk menyediakan *users* dengan suatu *drop-down list* dari *lookup* item untuk mengisi dalam *field*s yang membutuhkan data dari *dataset* lain.

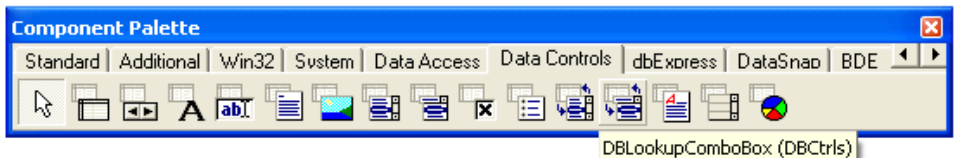

**Gambar 2.2 Objek DBLookupComboBox**

Secara umum kegunaan kedua objek ini adalah sama, namun dibedakan dengan tampilan atau fungsi dari objek dasar-nya, yaitu ComboBox atau ListBox. ComboBox dapat dibayangkan dengan sebuah kotak besar *drop-down* yang dapat digunakan untuk menampung sejumlah data. Sedangkan ListBox dapat dibayang seperti objek Edit yang disusun bertumpuk secara vertikal, dimana masing-masing tumpulan adalah index yang berisi data.

**Tabel 2.1**
**Properties Objek DBLookupComboBox/DBLookupListBox**

| Property | Nilai |
| --- | --- |
| DataSource | Mengacu pada sumber data (DataSource) dari form aktif |
| DataField | *Field* yang akan ditampilkan/disimpan ke dalam objek *database* |
| ListSource | Mengacu pada daftar sumber (ListSource) yang berasal dari sumber data (DataSource) milik Table Parent |
| ListField | Daftar *field* yang akan ditampilkan pada objek *database* yang mengacu pada Tabel Parent |
| KeyField | *Field* Kunci (Primary Key) yang ada pada Table Parent |
| Name | Nama objek *database* yang sedang digunakan |

### Borland Delphi

Borland Delphi merupakan Pemrograman Generasi IV yang memiliki kelebihan berupa tampilan visual yang *user-friendly*. Hal ini dikarenakan Borland Delphi dikembangkan dengan menggunakan Konsep *Object Oriented Programming* (OOP), sehingga memungkinkan penggunaan kembali (*re-use*) sejumlah objek/komponen untuk kepentingan aplikasi yang berbeda.

## 3. METODOLOGI PENELITIAN

### Objek Penelitian

Dalam kajian ilmiiah kali ini, objek penelitian bukanlah terhadap suatu aplikasi tertentu, namun dapat digunakan untuk beragam aplikasi yang menggunakan *lookup*. Perangkat lunak yang digunakan adalah Borland Delphi 7.0 produk dari Insprise Corp.

### Metode Pengumpulan Data

Dalam kajian ilmiah kali ini, penulis menggunakan kajian Studi Pustaka, yaitu dengan mempelajari literatur yang berhubungan dengan *lookup*, serta dengan Studi Laboratorium, yaitu dengan mencoba melakukan penggunaan objek dan/atau pengaturan *properties* pada objek *lookup* atau pada objek Tabel *Parent*. Adapun tabel yang penulis gunakan adalah tabel-tabel yang ada pada DBDEMOS, yaitu: 1) Customer.db, 2) Orders.db, 3) Items.db, 4) Parts.db, serta 5) Employee.db.

**Metode Pengembangan Sistem**

Setelah melakukan sejumlah pembuatan program, maka penulis berkesimpulan bahwa didapat sejumlah langkah-langkah yang dapat dilakukan secara sistematis dalam membangun suatu aplikasi visual yang menggunakan objek *lookup*. Adapun langkah tersebut adalah: 1) Rancangan Skema *Relational Database* (RD), 2) Rancangan Form, 3) Rancangan Skema Hubungan Antar Objek *Database*, 4) Pengaturan *Properties & Tools*, serta 5) Pembuatan Kode Program / *Procedures*.

**Rancangan Skema *Relational Database* (RD)**

Untuk memudahkan para pembaca dalam mencerna logika konsepsual *lookup*, ada baiknya sebelumnya membuat program (yang menggunakan *lookup*) Anda terlebih dahulu mengetahui/membuat relasi dari aplikasi yang akan dibuat. Dengan mengetahui gambaran relasi dari tabel yang terlibat dalam *database*, kita dapat dengan mudah menentukan tabel mana yang berfungsi sebagai Parent/Child. Sehingga kita dapat dengan mudah pula menentukan *source* dari tabel Parent yang akan di-*lookup*. Sebagai contoh lihat gambar 2.

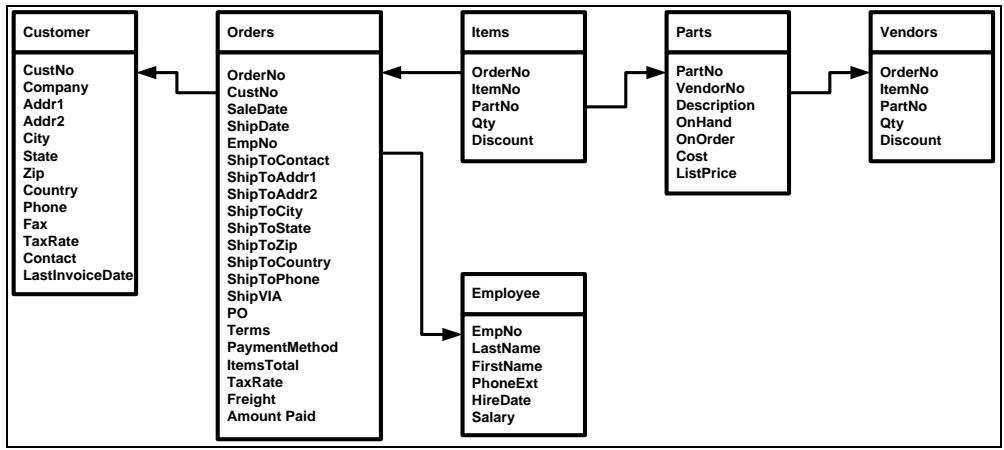

**Gambar 3.1 Skema *Relational Database* (RD)**

**Rancangan Form**

Setelah skema RD dari aplikasi yang akan dibuat telah diketahui, maka langkah selanjutnya, kita dapat mulai merancang form aplikasi yang diinginkan. Tempatkanlah objek.komponen yang dibutuhkan secara rapi, indah, menarik, efisien dalam pemanfaatan ruang (*space*) dari form. Sebagai contoh perhatikan rancangan form pada gambar 1.4. kita menggunakan kedua objek untuk *lookup*, yakni: DBLookupComboBox (untuk me-*lookup* tbCustomer) dan DBLoookupListBox (untuk me-*lookup* tbEmployee). Serta *lookup* dengan menggunakan pengaturan *Properties Field Type*, dengan menggunakan objek DBGrid.

**Gambar 3.2 Rancangan FormOrders**

**Rancangan Skema Hubungan Antar Objek Database**

Rancangan skema ini dilakukan untuk memudahkan logika dari *programmer* dalam mengatur *properties* dari masing-masing objek *database* yang berelasi. Sebagai contoh gunakan DataModule untuk memisahkan tempat perancangan objek-objek *database* yang non-visual.

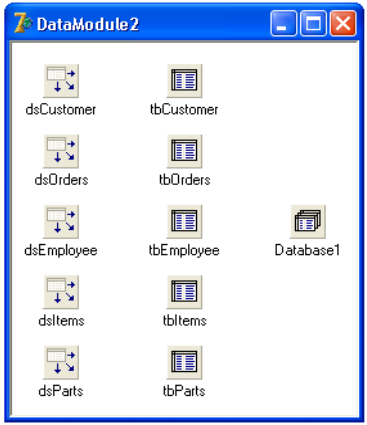

**Gambar 3.3 Skema Hubungan Antar Objek *Database***

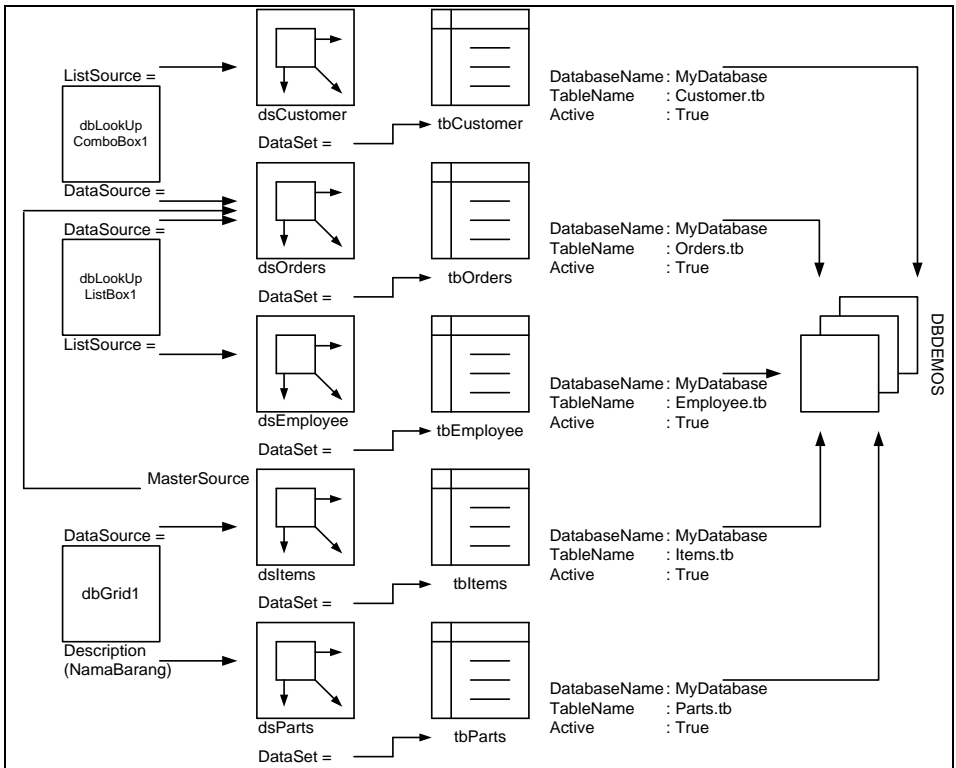

**Gambar 3.4 Skema Hubungan Antar Objek *Database***

**Pengaturan *Properties* dan/atau *Tools***

**Tabel 3.1**
**Pengaturan *Properties* Objek *Database* pada FormOrders**

| No. | Objek / Komponen | Property | Nilai | Event |
|---|---|---|---|---|
| 1. | Database1 | AliasName<br>DatabaseName<br>Name | DBDEMOS<br>MyDatabase<br>Database1 | |
| 2. | Table1 | DatabaseName<br>TableName<br>Active<br>Name | MyDatabase<br>Orders.tb<br>True<br>tbOrders | |
| 3. | DataSource1 | DataSet<br>Name | tbOrders<br>dsOrders | |
| 4. | DBEdit1 | DataSource<br>DataField<br>Name | dsOrders<br>OrderNo<br>dbeOrderNo | |
| | .<br>.<br>. | | | |
| 22. | DBEdit19 | DataSource<br>DataField<br>Name | dsOrders<br>AmountPaid<br>dbeAmountPaid | |
| 23. | DBGrid1 | DataSource<br>Name | dsOrders<br>DBGrid1 | |
| 24. | DBNavigator1 | DataSource<br>Name | dsOrders<br>DBNavigator1 | |
| 25. | Table2 | DatabaseName<br>TableName<br>Active<br>Name | MyDatabase<br>Customer.tb<br>True<br>tbCustomer | |
| 26. | DataSource2 | DataSet<br>Name | tbCustomer<br>dsCustomer | |
| 27. | DBLookupComboBox1 | DataSource<br>DataField<br>ListSource<br>ListField<br>KeyField<br>Name | dsOrders<br>CustNo<br>dsCustomer<br>CustNo;Company<br>CustNo<br>DBLookupComboBox1 | |
*Validasi Data dengan Menggunakan Objek Lookup … (Leon Andretti Abdillah)*     9

|  |  |  |  |
|---|---|---|---|
| 28. | Table3 | DatabaseName | MyDatabase |
| | | TableName | Employee.tb |
| | | Active | True |
| | | Name | tbEmployee |
| 29. | DataSource3 | DataSet | tbEmployee |
| | | Name | dsEmployee |
| 30. | DBLookupListBox1 | DataSource | dsOrders |
| | | DataField | EmpNo |
| | | ListSource | dsEmployee |
| | | ListField | EmpNo; FirstName; LastName |
| | | KeyField | EmpNo |
| | | Name | DBLookupListBox1 |
| 31. | Table4 | DatabaseName | MyDatabase |
| | | TableName | Items.tb |
| | | Active | True |
| | | MasterSource | dsOrders |
| | | MasterFields | OrderNo |
| | | Name | tbItems |
| 32. | DataSource4 | DataSet | tbItems |
| | | Name | dsItems |
| 33. | Table5 | DatabaseName | MyDatabase |
| | | TableName | Parts.tb |
| | | Active | True |
| | | Name | tbParts |
| 34. | DataSource5 | DataSet | tbParts |
| | | Name | dsParts |

Setelah semua *properties* yang dibutuhkan diatur dengan baik seperti pada tabel 1.1, maka sebelum kita memasukkan kode programnya, kita hendaklah memasukkan terlebih dahulu *fields* ke dalam objek tabel. Caranya, ikuti langkah-langkah sebagai berikut:

1) Pada objek tbOrders, klik kanan, → PopUpMenu,
2) Pada PopUpMenu, pilih Field Editor
3) Pada Field Editor, klik kanan → PopUpMenu,
4) Pada PopUpMenu, pilih Add all fields
5) Maka, pada Field Editor akan terisi dengan fields dari tbOrders.
6) Dengan cara yang sama tambahkan/masukkan semua fields dari tbItems, tbParts, tbCustomer, dan tbEmployee kedalam Field Editor masing-masing.

Hasil dari pengaturan Field Editor ini, dapat dilihat pada gambar 5. perhatikan semua Field Editor berisi *field*s dari tabel masing-masing. Dengan pengaturan ini, maka semua *field* yang ada dalam *table* dapat digunakan untuk beragam pengaturan.

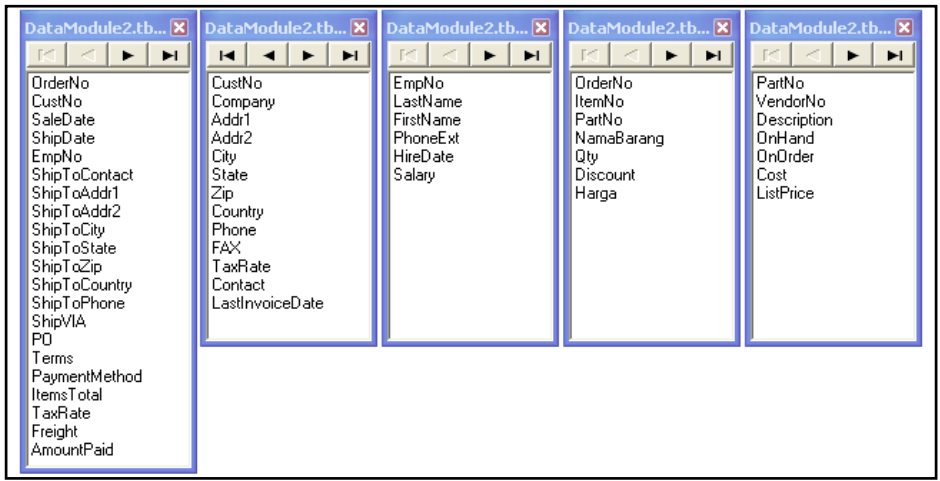

**Gambar 3.5 Field Editor dari tbOrders, tbCustomer, tbEmployee, tbItems, tbParts**

**Pemasukan Program Code/Procedures**

Setelah semua pengaturan *properties* dan tools dari objek-objek *database* yang terlibat dalam aplikasi selesai dilakukan, Anda dapat memasukkan kode program, sebagai berikut:

[1] **DataModule2Create**. Pada area DataModule2 kosong klik 2 (dua) kali. Kemudian masukkan kode program berikut:

```
procedure TDataModule2.DataModuleCreate(Sender: TObject);
begin
   tbOrders.Active   := True;
   tbItems.Active    := True;
   tbCustomer.Active := True;
   tbEmployee.Active := True;
   tbParts.Active    := True;
   (* Pengaturan DisplayFormat pada tbOrders *)
   tbOrdersTaxRate.DisplayFormat   := '#,#';
```

```
    tbOrdersItemsTotal.DisplayFormat := 'US$ #.#,#';
    tbOrdersFreight.DisplayFormat    := 'US$ #.#,#';
    tbOrdersAmountpaid.DisplayFormat := 'US$#.#,#';
    (* Pengaturan DisplayFormat pada tbItems *)
    tbItemsQty.DisplayFormat         := '#.#';
    tbItemsDiscount.DisplayFormat    := '#,#';
    tbItemsHarga.DisplayFormat       := 'US$ #.#,#';
end;
```

[2]  **tbItemsCalc*Fields***. Ikuti pengaturan *properties*-nya sebagai berikut:

Adapun langkah-langkah dalam membuat tbItemsCalcFields, adalah : 1) Pada tbItems dalam DataModule2 klik 1 (satu) kali, 2) Pada Object Inspector klik Tab Event, 3) Pada Tab Events pilih OnCalcFields, 4) Dan pada ComboBox di sebelah kanannya klik 2 (dua) kali, dan 5) Masukkan kode program sebagai berikut :

```
procedure TDataModule2.tbItemsCalcFields(DataSet: TDataSet);
var Jumlah : Integer;
    Harga, Diskon, TotalHarga : Variant;
begin
    (* Mengambil nilai dari tbPartsListPrice *)
    Harga      := tbPartsListPrice.Value;
    (* Mengambil nilai dari tbItemsQty dan tbItemsDiscount *)
    Jumlah     := tbItemsQty.Value;
    Diskon     := tbItemsDiscount.Value/100;
    (* Proses perhitungan TotalHarga *)
    TotalHarga := ((1 – Diskon) * Harga) * Jumlah;
    (* Menampilkan TotalHarga pada tbItemsHarga *)
    tbItemsHarga.Value := TotalHarga;
end;
```

Simpan = F2, Run = F9.

## 4. PAMBAHASAN

### Lookup EmpNo

Pada saat program dijalankan, maka secara otomatis pointer akan dibawa ke posisi *record* dengan index 0, yaitu OrderNo = 1003. Perhatikan objek DBLookupListBox (sudut kiri bawah form) yang menampilkan data *employee*, sesuai dengan hasil pengaturan *properties*-nya. Dengan cara ini maka pemasukan data yang berhubungan dengan data *employee* hanya akan mengambil dari table *employee*.

Selain itu, operator tidak akan membuat kesalahan dalam memasukkan data *employee*. Kelebihan lain dari objek DBLookupListBox, adalah dapat menampung sejumlah (semua) *fields* yang ada dari *table*-nya (employee.db). Namun untuk contoh kali ini, kita hanya menggunakan 3 (tiga) *field*s saja, yaitu : EmpNo, FirstName, dan LastName. (Lihat gambar 6). Walaupun yang ditampilkan ada 3 (tiga) *field*s, tetapi data yang akan disimpan hanya EmpNo saja.

**Gambar 4.1 Objek Lookup DBLookupListBox, utnuk EmpNo**

**Lookup CustNo**

Pada gambar 7, coba Anda perhatikan objek DBLookupComboBox. Ia berisi data-data dari table customer.db. sehingga *customer* yang di*input*kan adalah yang valid. Dalam contoh yang diberikan CustNo yang diambil adalah CustNo 1351. Agar tidak terjadi kesalahan dalam pengakses data yang diinginkan, maka ada baiknya kita tambahkan kolom Contact. Penambahan ini berguna untuk memastikan bahwa data yang dimasukkan adalah yang memenuhi transaksi. Namun *database* hanya akan menyimpan data CustNo saja.

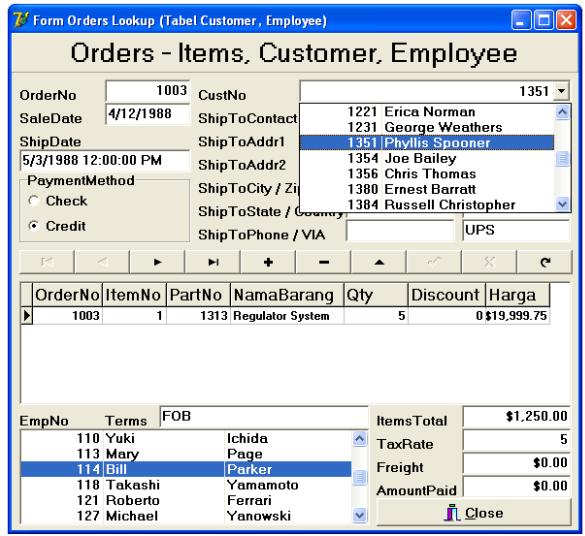

**Gambar 4.2 Objek Lookup DBLookupComboBox, untuk CustNo**

Lookup PartNo

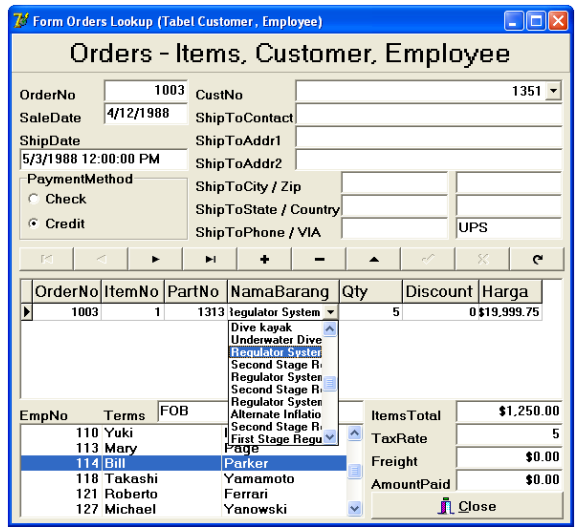

**Gambar 4.3 DBGrid Menampilkan *Lookup* NamaBarang**

Pada gambar 8, coba Anda perhatikan objek *Lookup* pada DBGrid. Ia tidak hanya berisi berisi data-data dari tabel part.db, namun juga sejumlah data darai tabel Order, Item, dan Part. Penggunaan *lookup* dapat memastikan bahwa suatu *description* yang di*input*kan adalah yang valid. Dalam contoh yang diberikan Description yang diambil adalah Description dari PartNo = 1313. berbeda dengan objek DBComboBox, suatu kolom dalam DBGrid hanya mampu menampilkan 1 (kolom) dari *field* tertentu. Penambahan NamaBarang tidak akan memperbesar kapasistas penyimpan, karena ia hanya digunakan untuk memastikan bahwa barang yang diinginkan adalah valid.

5. **KESIMPULAN**

Setelah melakukan percobaan terhadap sejumlah studi kasus (yang menggunakan *lookup*), maka dapat diambil sejumlah kesimpulan, yakni :
1. Validitas input atas data sangat penting, untuk mendukung integrasi antar table yang berelasi dalam suatu aplikasi.
2. *Lookup* dapat menggunakan pengaturan *properties* pada objek/komponen DBLookupComboBox/DBLookupListBox, atau dengan cara Pengaturan *properties* pada objek Tabel.
3. Objek / Komponen DBLookupComboBox / DBLookupListBox, dapat digunakan untuk mewakili Tabel *Parent* yang akan di-*lookup* / diambil datanya pada Tabel *Child*.
4. Tingkat akurasi input data sangat terjamin validitasnya dengan ke-2 cara / kedua objek tersebut.
5. Dapat digunakan untuk mewakili relasi 1-1, 1-N/M-1, atau M-N

# DAFTAR RUJUKAN